\documentclass[twocolumn]{article}

\pdfoutput=1

\usepackage{fullpage}
\usepackage[hyperindex,breaklinks]{hyperref}

\newtheorem{problem}{Problem}[section]
\newtheorem{definition}{Definition}[section]
\newtheorem{question}{Question}[section]

\title{Ark: A Real-World Consensus Implementation}
\author{Zardosht Kasheff \\
  \url{zardosht@tokutek.com} \and
  Leif Walsh \\
  \url{leif@tokutek.com}}
\date{}

\begin{document}
\maketitle

\begin{abstract}
  Ark is an implementation of a consensus algorithm similar to
  Paxos\cite{paxos} and Raft\cite{ongaro_search_2013}, designed as an
  improvement over the existing consensus algorithm used by
  MongoDB\textsuperscript{\textregistered} and TokuMX\texttrademark.

  Ark was designed from first principles, improving on the election
  algorithm used by TokuMX, to fix deficiencies in MongoDB's consensus
  algorithms that can cause data loss.  It ultimately has many
  similarities with Raft, but diverges in a few ways, mainly to support
  other features like chained replication and unacknowledged writes.
\end{abstract}

\section{Introduction}
\label{intro}

Databases with built-in replication are ubiquitous.  A short,
non-exhaustive list of such databases and database-like technologies is
Cassandra, Couchbase, CouchDB, FoundationDB, Kafka, Microsoft SQL Server,
MongoDB, Oracle SQL, PostgreSQL, RabbitMQ, Redis, Riak, Zookeeper.  Since
in many use cases, replicas are treated as insurance against hardware and
software faults, replication must provide strong, clear, and correct
guarantees to the application layer.  As such, there is a great deal of
interest recently in understanding how these technologies work in theory
and in practice.

MongoDB is a NoSQL DBMS with built-in leader/follower asynchronous
replication.  MongoDB's Write Concern API\cite{mongodb:write_concern}
allows clients to choose from a spectrum of write consistency ranging from
entirely unsafe (``unacknowledged'') through fully consistent
(``majority'').

TokuMX\footnote{TokuMX is a fork of MongoDB developed and maintained by
  Tokutek, Inc. (\url{http://www.tokutek.com}).} is a fork of MongoDB with
a range of improvements centered on a different storage engine based on
Fractal Tree indexes\textsuperscript{\textregistered}\cite{BenderFaFi07}.

MongoDB is known to have problems with its replication
algorithm\cite{jepsen:mongodb}.  In short, the semantics offered by the
Write Concern API are not reliably delivered in the face of network
failures, which has contributed to a widespread sense of distrust in
MongoDB.

In TokuMX, we have changed the replication election protocol to deliver
the semantics of the Write Concern API in a provably correct way.  In
particular, this means that for clients using the ``majority'' Write
Concern, TokuMX is a proper ``CP'' system, in the sense of Brewer's
so-called ``CAP Theorem''\cite{Brewer:2000:TRD:343477.343502}.

As Ark is primarily a modification of the existing MongoDB consensus
algorithm, it's important to start with an understanding of that.  In
Section~\ref{mongo-repl}, we explain the goals and relevant aspects of the
existing replication algorithm in MongoDB 2.6 and TokuMX 1.5.  In
Section~\ref{problems}, we detail how the current replication algorithm
fails to deliver on the CP promise of the ``majority'' Write Concern, and
causes two other related problems.

We describe our changes to the replication algorithm in
Section~\ref{ark-design}, turning it into a new algorithm called Ark, and
in Section~\ref{solution-raft}, we draw parallels to the Raft paper and
its safety and liveness proofs.  We discuss specific implementation
choices and tradeoffs in Section~\ref{future-work}.

\section{MongoDB replication}
\label{mongo-repl}

In MongoDB/TokuMX, the purpose of replication is to have multiple
machines, or replicas, store the exact same data.  This is leader/follower
replication, as opposed to multi-master, which means that there should be
a totally ordered sequence of client operations that is identical on every
replica in the set.  It is asynchronous, which means that followers pull
updates from the leader and announce their synced position in the
operation history, rather than the leader pushing changes to followers and
waiting synchronously for them to respond.

A ``replica set'' is composed of a leader (``primary'') and a set of
followers (``secondaries'').  Replication's main goals are to synchronize
the data on each member of the set (``replica''), to provide an API for
clients to understand how much their updates have been replicated through
the set, and to automatically fail over to a new primary if the primary
crashes or is disconnected from the set.

\subsection{Synchronization}
\label{mongo-sync}

Synchronization within the replica set is accomplished as follows:

One replica in the set is designated as primary, or the leader.  The
primary is the only replica that accepts updates from clients.  All other
replicas in the set are secondaries, or followers.

Details of each client update are written to the ``oplog'', a totally
ordered sequence of operations (similar to the binary log in MySQL, can
also be thought of as the replication log).  Entries in the oplog have a
position associated with them, that defines the order of updates.
Secondaries maintain an identical copy of the oplog, and apply
modifications in the order they appear in the oplog.

Secondaries cannot be modified by users.  The job of a secondary is to
constantly pull data from the primary's oplog, save it in its own oplog,
and apply the operations to its copies of the user data collections.  A
secondary may serve some client reads from its data collections, or simply
be available to be elected to primary if failover is necessary.

An important point here: the secondary does not have to be pulling data
directly from the primary.  Any secondary is allowed to pull data from any
other replica in the replica set, as long as the other replica's
position is ahead of the secondary's position.

Every two seconds, all replicas exchange information with each other, in
what is called ``heartbeats''.  This information can be arbitrary.  The
two most important pieces of information exchanged in the current
algorithm are the replica's current oplog position and the fact that a
replica is still up and responding.  One thing that causes a replica to
decide it is time to try to fail over to a new primary is the failure to
receive a heartbeat from the current primary.

\subsection{Write Concern}
\label{mongo-write-concern}

The Write Concern API provides a mechanism by which clients may block
until $0$, $1$, or more replicas have acknowledged an update.  A Write
Concern of ``majority'' blocks until a majority of replicas ($\lfloor n/2
\rfloor + 1$, if the replica set contains $n$ replicas) acknowledge the
update.

The purpose of ``majority'' Write Concern is to make the replica set fully
consistent.  Since a node can only be elected primary by a majority (see
Section~\ref{mongo-failover}), if each update is only considered finished
once acknowledged by a majority, any successful election must include a
replica which was aware of the acknowledged update, and it will be that
replica's responsibility to ensure that the update persists.

Clients can use ``majority'' Write Concern on every update, which
guarantees that the update will be persisted even if failover happens
immediately after the update succeeds and reports majority
acknowledgement.

Note that false negatives can still occur in this asynchronous model: an
update may be replicated to a majority of replicas and be safe, but the
acknowledgement might fail to return to the client.  In Ark, just as in
Raft, we consider this an acceptable failure mode.  Applications that
require stronger guarantees than this should use a system that implements
a distributed atomic commit protocol like
2PC\cite{Gray:1978:NDB:647433.723863,Lampson:1981:AT:647434.726386}, for
example, Apache Zookeeper.  Note that the ``clients'' in the Raft paper
are similar to the database engine in TokuMX with Ark.  In TokuMX, entries
are committed to the oplog in the same transaction as they are applied to
collections (the state machine in Raft), so the false negatives here do
not result in multiple delivery of oplog messages, rather our false
negative is at the higher database API layer.

We consider ``majority'' Write Concern to be the most interesting use case
to support, since any weaker Write Concern expresses the client's
willingness to suffer data loss in some scenarios.  The primary goal of
Ark is to provide correct ``majority'' Write Concern, but we have also
made some choices that make it more suitable for weaker consistency
choices that are useful in some deployments.

\subsection{Failover}
\label{mongo-failover}

Failover is the mechanism by which the replica set attempts to ensure that
there is always exactly one primary available to serve client updates.  If
the primary crashes, is partitioned away from the majority of replicas by
a network failure, or is otherwise unresponsive, the replica set attempts
to select another replica to step up as primary.

A replica is deemed unreachable if a heartbeat request fails.  At a high
level, if the primary becomes unreachable by other replicas of the replica
set, the other replicas will try to hold an ``election'', a process by
which they elect a new primary to start accepting user updates.  A
majority ($\lfloor n/2 \rfloor + 1$) of replicas in the replica set are
required to elect a new primary.

Note that the election process and the data synchronization process (the
threads on a secondary responsible for picking a replica to copy data
from) are fairly independent.  There is a little synchronization used to
determine oplog position, but that is it.  It’s possible for a secondary
to be replicating oplog data from a replica that the majority component of
replicas thinks is unreachable.

An important component of failover is rollback.  Consider the following
scenario:
\begin{itemize}
\item Replica $A$ (the primary) receives an update, which it applies and
  logs.  A network partition isolates $A$, a failover happens, and new
  primary, $B$, is elected before that update is replicated.
\item $B$ does not have the oplog entry for this update.  In fact, by
  electing $B$, all of the replicas that elected $B$ agreed that the oplog
  does not contain this update.
\item At some point, when $A$ gets reconnected to the set, it will see
  that the current primary $B$ does not have this update in its oplog, and
  will ``rollback'', or undo, the update.  This is similar to the way a
  leader in Raft forces its followers to replace log entries that differ
  from its own.

  For simplicity, assume all operations that appear in the oplog can be
  reversed.
\end{itemize}

So, in summary, there are three important concepts there:
\begin{enumerate}
\item Data synchronization.
\item Write acknowledgement.
\item Elections/failover and rollback.
\end{enumerate}

The property we wish to have is the following.  If a client successfully
gets an acknowledgement for ``majority'' Write Concern for some update,
then the user knows that the update is guaranteed to be in the replica set
going forward, and will not be rolled back.

\subsection{TokuMX differences}
\label{toku-repl-differences}

The replication concepts in TokuMX 1.5 are largely the same as in MongoDB
2.6.  However, there are a few minor differences that are worth mentioning
before we continue.

\subsubsection{GTID}
\label{toku-gtid}

MongoDB's oplog entries are ordered by a data type called \verb+OpTime+,
which is the concatenation of a 32-bit timestamp with second resolution,
and a 32-bit counter that is incremented with each operation, and reset to
$1$ each time the timestamp increases.

To support better concurrency, and in anticipation of some of the changes
described in this report, TokuMX has always used a different data type to
order oplog entries, called the \verb+GTID+.  This is a pair of 64-bit
integers, $\langle term, opid \rangle$, where the $opid$ is incremented
each time an operation commits on the primary, and the $term$ is
incremented each time a new primary is elected, and at this point the
$opid$ is reset to $0$.  The \verb+GTID+ is compared lexicographically, so
$\langle 3, 100 \rangle < \langle 3, 101 \rangle < \langle 4, 0 \rangle$.

\subsubsection{Idempotency}
\label{toku-idempotency}

MongoDB's oplog entries are required to be idempotent.  This permits the
oplog application to support at-least-once delivery, but limits the range
of operations that can be expressed in the oplog.

TokuMX's oplog entries are not required to be idempotent, therefore the
oplog entries must be applied exactly once on secondaries.  Given that the
secondaries are already designed to store exact copies of the oplog, this
is simple to implement by just locally noting which entries have and have
not been applied.  This is functionally equivalent to the suggestion in
Raft that clients assign serial numbers to commands, and have the state
machine avoid re-executing commands.

This removal of the idempotency restriction is to implement other
optimizations in the application of oplog entries, and is mostly
inconsequential to the problem of consensus.  The point of consensus in
the replication system is to copy the oplog perfectly, and application is
effectively orthogonal.

\subsection{Failover details}
\label{failover-details}

TokuMX’s current election/failover protocol, inherited from MongoDB, is as
follows.  Suppose there is a network partition such that the primary $A$
is disconnected from the replica set.  There are two independent things
that must happen:
\begin{enumerate}
\item Another replica $B$ notices that it cannot reach $A$, looks at what
  it knows of the state of the replica set (via heartbeat messages it has
  received), and decides ``I think I'll make a good primary''.  $B$ then
  proceeds to elect itself as the set's new primary.
\item Independently, $A$ notices that it cannot reach a majority of the
  set, and decides to transition from primary $\rightarrow$ secondary.

  The consequence of this is that $A$ will stop accepting update
  operations from clients, because it is no longer convinced that a
  majority of replicas will acknowledge those updates.
\end{enumerate}

Once $B$ decides to elect itself, it does so with the following procedure:
\begin{itemize}
\item $B$ broadcasts a message to all other replicas asking ``should I try
  to elect myself?''.  This is known as a {\bf speculative election} ({\em
    this is for heuristic purposes}).  With the replies, $B$ learns:
  \begin{itemize}
  \item Whether anyone would veto the election.

    If so, $B$ does not try to elect itself.
  \item Whether any replica has an oplog that is ahead of $B$'s oplog.

    If so, $B$ does not try to elect itself.
  \item If $B$ is at the same oplog position as another potential primary.

    If so, it will sleep for a random period of time between 50 and
    1050ms, and then try the election protocol again.  The next time it
    tries, it will remember that it just slept and not sleep again.

    The purpose of this step is if two replicas may both be eligible to
    become primary, a random sleep will help ensure that concurrent
    elections are not run.
  \end{itemize}
\item If all of the respondents say ``yes'', then it proceeds with the
  {\bf authoritative election}.  $B$ broadcasts a message that states ``I
  wish to elect myself as primary''.

  The protocol has begun.  When each replica gets this message, it does the
  following:
  \begin{itemize}
  \item If there is a reason to veto, then it replies ``veto'' (e.g. $B$'s
    understanding of the set membership is stale, or another primary
    exists).

    A ``veto'' vote is functionally equivalent to ``no'', in fact the
    election protocol works properly if ``veto'' is interpreted as ``no'',
    but ``veto'' is an optimization replicas can use when they have
    concrete evidence that the election should not succeed.
  \item If the replica participating in the election has voted ``yes'' for
    anyone in the last 30 seconds (in an authoritative election; votes
    cast in a speculative election are not considered), then vote ``no''.

    A replica may vote ``yes'' (in an authoritative election) only once
    every 30 seconds.
  \item Otherwise, vote ``yes''.  

    The replica doesn't bother looking at the oplog's position before
    voting yes.  This last fact is strange, but true, which Zardosht
    pointed out in
    \url{https://groups.google.com/forum/?hl=en-US\&fromgroups#!topic/mongodb-dev/lH3hs8h7NrE}.
  \end{itemize}
\item If a majority reply to $B$ saying ``yes'', and no replica responds
  ``veto'', then $B$ declares itself as primary and begins accepting
  client updates.
\end{itemize}

If at any point, a primary notices that another primary exists, it blindly
steps down.  The hope is to have another election resolve the dual primary
issue.  If $A$ and $B$ are both primary, when they exchange heartbeats,
they will step down.

\section{Problems}
\label{problems}

Briefly, TokuMX and MongoDB currently have the following problem: a user
may successfully get an acknowledgement for ``majority'' Write Concern for
some update, and that update may not survive.  Despite the fact that the
update was acknowledged by a majority of replicas, the update may later
roll back.

At a high level, there are three problems with the election protocol, and
they are all loosely related:
\begin{enumerate}
\item Updates that succeed with ``majority'' Write Concern may roll back.
  Otherwise known as ``data loss''.  This is the big problem mentioned
  above.
\item Multiple primaries don't resolve themselves in an intelligent way.
\item Letting a replica vote once every 30 seconds can lead to some long
  failover times if elections make it to the second phase and fail.
\end{enumerate}

Now let's dig into each of these problems a bit more.

\begin{problem}
  \label{problem-rollback-majority}
  Write Concern of ``majority'' may not prevent the rollback of updates.
\end{problem}

Before discussing the problems with this protocol, let's first say what is
not a problem: having $A$ accept updates (temporarily) after becoming
disconnected is not a problem.  When $A$ reconnects with the replica set,
these additional updates will be rolled back.  But this is okay, as long
as we haven't yet guaranteed to any clients that those updates are safe.

Instead, the problem with this protocol is the following: updates may get
an acknowledgement of ``majority'' Write Concern (which does mean they're
safe), and still be rolled back.

Consider the following scenario:
\begin{itemize}
\item A network partition happens such that $A$ is disconnected from the
  replica set.
\item The rest of the set elects $B$ as primary, but $A$ has yet to
  transition to secondary (because that process is independent of the rest
  of the replica set's election, in fact $A$ may not yet know it has been
  disconnected).
\item $A$ then reconnects with the replica set, thinking it is still
  primary.
\item Now we have two primaries, $A$ and $B$.  Also, recall that all other
  replicas in the set may be syncing the oplog from any other replica.
  They may all be replicating off $A$ or $B$.

  This is the big problem.  An update may replicate off of $A$ or $B$, and
  get acknowledged by a majority of the replica set.
\item At best, $A$ and $B$ realize they are both primary, step down, and
  allow an election to take place.

  In Problem~\ref{problem-multiple-primaries}, we state why it can be
  worse.
\end{itemize}

Now we have a problem.  Either $A$ or $B$ can win the new election, and
the loser may have some update that was acknowledged by a majority of the
replica set.  This update will then be rolled back, violating the
``majority'' Write Concern contract with the client.

The fundamental flaw in the current replication algorithm that leads to
this behavior is that secondaries blindly acknowledge any update that they
can copy, regardless of whether the update may later rollback.  If a
secondary acknowledges an update from $A$ after having voted $B$ as
primary, there is something wrong.  The secondary should know that $A$'s
new updates may be rolled back and therefore must not be acknowledged.

\begin{problem}
  \label{problem-multiple-primaries}
  Multiple primaries don't resolve themselves deterministically.
\end{problem}

This is related to the first problem.  We want multiple primaries to
resolve themselves in a predictable way because we want to be able to
predict which updates will survive and which will be rolled back.

Right now, if two primaries exist, and are made aware of each other via
heartbeats, then they step down and let another election take care of the
problem.  This is problematic for the following reasons:
\begin{enumerate}
\item It requires two primaries seeing each other to resolve the issue.
  In the right kind of network partition, this may take an indefinitely
  long time.  See \url{https://jira.mongodb.org/browse/SERVER-9848}.
\item If one primary gets a heartbeat message before it can send one to
  the other, that one primary will step down and the other will not.  This
  makes ``which primary wins'' essentially arbitrary.  Having any decision
  here be arbitrary is bad, because it makes the future unpredictable.
\end{enumerate}

\begin{problem}
  \label{problem-long-elections}
  Elections may take a long time because a member can vote ``yes'' at most
  once every 30 seconds.
\end{problem}

MongoDB's election protocol requires that a member may not vote ``yes'' in
more than one election in any 30-second period.
We think this is because the order of oplog entries is primarily
determined by the timestamp on the primary where they are created, so if
successful elections happen too frequently, updates to successive
legitimate primaries may end up getting reordered in the oplog as a result
of clock skew.  This would mean that, logically at least, there would have
been multiple concurrent primaries, and the MongoDB replication system is
not equipped to properly resolve this situation.  Limiting elections to
succeed at most once every 30 seconds means that if the maximum clock skew
among replicas is less than 30 seconds, the order of updates done by a
single client will not be changed too much by elections.  Updates from one
primary may be interspersed with the successive primary's updates, but not
with the one after that.

However, this 30 second threshold can be problematic in practice,
especially if an election fails: this necessarily makes the set
unavailable for at least 30 seconds, maybe more if successive elections
fail.

Part of the problem in this case is that the candidate does not do a good
job of using the first phase of the election process to determine whether
the second phase will succeed.  That is bad, because a failed second phase
can really elongate the downtime during a failover.  Basically, if the
first phase can determine that the second phase will be unsuccessful, it
should do so.  Currently, MongoDB and TokuMX have the following two
issues:
\begin{enumerate}
\item \url{https://jira.mongodb.org/browse/SERVER-14382}: If a member will
  vote ``no'' in the second phase because it has voted ``yes'' for someone
  else in the last 30 seconds, it should notify the candidate of this in
  the first phase.
\item \url{https://jira.mongodb.org/browse/SERVER-14531}: If the candidate
  gets responses from less than a majority of replicas during the first
  phase, it should not proceed onto the second (authoritative) phase.
\end{enumerate}

\section{Ark Design}
\label{ark-design}

At a high level, we want the solution we employ to have the following
characteristics:
\begin{itemize}
\item Any update that succeeds with majority Write Concern cannot be
  rolled back during a failover.
  (Problem~\ref{problem-rollback-majority})
\item Multiple primaries should resolve themselves in a predictable way.
  (Problem~\ref{problem-multiple-primaries})
\item Failover times should be faster.  We want to get rid of this 30
  second timer between elections.  (Problem~\ref{problem-long-elections})
\end{itemize}

We'll start by restricting ourselves to the TokuMX replication algorithm,
namely that oplog entries are identified by and ordered according to the
\verb+GTID+, as defined in Section~\ref{toku-gtid} to be $\langle term,
opid \rangle$.

The key difference between Ark and the standard MongoDB replication
algorithm is the $term$ part of a \verb+GTID+.  This will be used to
demarcate elections and to provide an association between election terms
and client update operations.  This association is critical for ensuring
that updates with ``majority'' Write Concern cannot be rolled back.

Below, we will consider an election in a replica set led by a primary $A$.
In the voting protocol, another replica $B$ tries to elect itself the new
primary.

\subsection{Election changes}
\label{election-changes}

The first protocol change is to associate the $term$ in the oplog's
\verb+GTID+ with elections.  Each authoritative election is identified
with an $electionTermId$.  Every replica in the set considers
$electionTermId$ to be a strictly monotonically increasing sequence.

\begin{definition}[$electionTermId$]
  Each authoritative election is identified by an $electionTermId$ that is
  selected by the candidate $B$.

  If elected, this replica must use the $electionTermId$ from its
  successful election as the $term$ part of the \verb+GTID+s it creates
  for new oplog entries during its tenure.
\end{definition}

The second protocol change is that each replica maintains a local value
$maxVotedTermId$, which is always greater than or equal to the $term$ in
any \verb+GTID+ in its oplog.

\begin{definition}[$maxVotedTermId$]
  A replica's $maxVotedTermId$ is the maximum $electionTermId$ for any
  election in which it voted ``yes''.
\end{definition}

A replica's $maxVotedTermId$ will be used to decide which updates to
acknowledge and which elections to participate in.

During normal operation, this will be identical to the $term$ part of the
\verb+GTID+ for elements being added to the oplog by the current primary.

\subsubsection{Term ID maintenance}
\label{term-id-maintenance}

These two values are propogated and maintained through a few different
mechanisms:

\begin{itemize}
\item In speculative elections, when a replica votes ``yes'', ``no'', or
  ``veto'', it includes its $maxVotedTermId$ along with its ballot.

  In the speculative election, the candidate learns about each other
  responding replica's $maxVotedTermId$.  If the candidate decides to
  proceed with an authoritative election, it specifies
  \[
  electionTermId = \max_{ballots} maxVotedTermId + 1
  \]
  and sends this $electionTermId$ along with its authoritative election
  request to all voters.
\item In authoritative elections, when a replica votes ``yes'' in an
  election, it updates its $maxVotedTermId$ to the election's
  $electionTermId$.

  Note that this $maxVotedTermId$ is persistent to disk and survives
  crashes and process restarts.
\end{itemize}

\subsubsection{Voting changes}
\label{voting-changes}

When voting in an authoritative elections, each replica considers the same
conditions as in the old algorithm, except that we eliminate the
restriction against voting ``yes'' in two elections in the same 30-second
time span, and add a few new conditions:
\begin{enumerate}
\item If $electionTermId \le maxVotedTermId$, that is, the voter has
  already voted ``yes'' in another election with the same or higher
  $electionTermId$, it votes ``no''.

  This makes sure that each replica may only vote for one candidate per
  $electionTermId$.
\item If the last \verb+GTID+ in the candidate's oplog is less than some
  \verb+GTID+ in the voter's oplog, the voter responds with ``veto''.

  {\em Note that this check was not done in the old algorithm.}  It is not
  necessary to veto at this point, a ``no'' vote would be sufficient to
  support ``majority'' Write Concern properly, but we will discuss this
  choice later, in Section~\ref{future-work}.
\end{enumerate}

Since a replica only votes ``yes'' in elections where $electionTermId >
maxVotedTermId$, and it updates $maxVotedTermId$ when voting ``yes'', this
establishes that the sequence of $maxVotedTermId$ values for each replica
in the set is strictly monotonically increasing.

Additionally, a successful election requires a majority of ``yes'' votes,
which implies that in every adjacent pair of successful elections there is
at least one ``yes'' voter in common.  Therefore, the replica set's total
sequence of successful $electionTermId$s (and therefore the sequence of
$term$s in \verb+GTID+s in the oplog) is also strictly monotonically
increasing.

\subsubsection{Election changes' effects}
\label{election-changes-effects}

These changes are enough to solve the problems of multiple primaries not
resolving themselves intelligently
(Problem~\ref{problem-multiple-primaries}), and of elections possibly
taking a long time due to the 30 second rule
(Problem~\ref{problem-long-elections}).  However, they are not yet enough
to prevent the primary problem that ``majority'' Write Concern is not
enough to prevent rollback (Problem~\ref{problem-rollback-majority}).

\subsection{Write Concern changes}
\label{write-concern-changes}

To solve Problem~\ref{problem-rollback-majority}, we make one more simple
change: a replica never {\em acknowledges} an update from a primary after
it votes ``yes'' in a later election.

Recall that once a replica votes ``yes'' in an election it updates its
$maxVotedTermId$ to that election's $electionTermId$.  The
$electionTermId$ must be greater than any $term$ in the voter's oplog, or
it would not have voted ``yes''.

This establishes a relationship between elections and Write Concern
acknowledgement, through the \verb+GTID+s in the oplog.  Namely, if a
replica acknowledges an update, it may not vote ``yes'' for a candidate
that doesn't have that update, and once a replica has voted ``yes'' for a
new primary, it will not acknowledge any more updates from older
primaries, so it cannot acknowledge updates that might later be rolled
back if that election succeeds.

Note that a replica may still copy and apply oplog entries from any member
of the set, knowing that it may roll them back at any point in the future.
This restriction only affects {\em acknowledgement}, and it is
fundamentally a simple rule: a replica may only acknowledge updates that
it thinks will not be rolled~back.

With this, any successful election will cause a majority of replicas to
cease acknowledgement of new updates to the old primary, which, if the
client uses ``majority'' Write Concern, means that successful updates must
have happened before those replicas voted for a new primary.  Therefore
those updates must be on the newly elected primary, and will not be rolled
back.  Updates with a weaker Write Concern may still be rolled back, but
this is a design principle we accept, in accordance with the Write Concern
design.

\subsubsection{Primary step-down}
\label{primary-step-down}

This new change introduces a possible problem: after a failed election,
all the replicas that voted ``yes'' will be ineligible to satisfy Write
Concern from the still-legitimate primary.  Over multiple failed
elections, this could render a majority of the set ineligible to
acknowledge updates, which would halt a system using ``majority'' Write
Concern (no update could be satisfactorily acknowledged).

The fix for this is to make the primary more sensitive: each heartbeat
message contains a node's $maxVotedTermId$, and when a primary $A$ sees a
heartbeat with $maxVotedTermId$ greater than the $electionTermId$ for the
election in which it was elected, the primary immediately steps down.  It
does this because it knows that some nodes in the set are not
acknowledging its updates (because they've voted ``yes'' in a later
election).  If $A$ should continue as primary, it may of course re-elect
itself immediately, but it will not continue accepting updates with
\verb+GTID+s that may not be acknowledged.

Once an election succeeds, the old primary will be in one of two cases:
\begin{itemize}
\item It may still see a majority of the set, in which case it will
  shortly receive a heartbeat containing a newer $maxVotedTermId$ and step
  down.
\item It may not see a majority of the set, in which case it will
  determine that it is out of touch with the majority component and will
  step down.
\end{itemize}

Thus, a deposed primary will necessarily step down after at most one
heartbeat period plus the heartbeat timeout period (which as of now totals
to about 12 seconds, and is configurable).

It is sufficient, to cause an old primary to step down, to have each
replica broadcast only the $maxVotedTermId$ for elections in which it
actually voted ``yes''.  However, to speed things up, we add some
optimizations.

\begin{definition}[maxKnownTermId]
  A replica's $maxKnownTermId$ is the maximum $electionTermId$ for any
  election it knows of where at least one replica voted ``yes''.
\end{definition}

Because a replica knows about itself,
\[
maxKnownTermId \ge maxVotedTermId
\]
no matter what.

Every heartbeat carries with it the node's $maxKnownTermId$.  If a replica
receives a heartbeat with a higher $maxKnownTermId$ than its own, it
replaces its own with the higher value, and in the future will broadcast
and use that value.

It is an optimization for replicas to rebroadcast their neighbors' higher
$maxKnownTermId$ value to attempt to propogate the news of an election
faster through a partially connected set.  A primary will then step down
if it sees a heartbeat with $maxKnownTermId$ greater than its
$electionTermId$.

An additional optimization is that, upon successful election, the newly
elected primary will ask all replicas to immediately send a heartbeat
broadcasting this news, rather than waiting for the heartbeat timer to
announce it.

\subsection{Addressing the problems}

Let's take a moment to look back at the original problems we introduced in
Section~\ref{problems} and reiterate how they are addressed.

\begin{problem}
  Write Concern of ``majority'' may not prevent the rollback of updates.
\end{problem}

No single machine will elect two different primaries with the same
$electionTermId$.  Because a majority is needed to elect a primary, any
two successful elections must have one voter in common, that voted ``yes''
in both elections.  Because no replica ever votes ``yes'' in two different
elections with the same $electionTermId$, no two successful elections can
have the same $electionTermId$.

Each voter that acknowledges an update will not vote ``yes'' in an
election that will rollback the update.  A majority that acknowledge an
update must include at least one member in the majority that vote ``yes''
in a subsequent election.  Therefore, if an update is acknowledged by a
majority, it must be on the new primary and so won't be rolled back.

For the contrapositive, a replica that votes ``yes'' to elect a new
primary will stop acknowledging updates from the old primary.  Therefore,
in a successful election, a majority of replicas will agree to stop
acknowledging updates from the old primary, and this will ensure that
those updates cannot have had ``majority'' Write Concern satisfied.

\begin{problem}
  Multiple primaries don't resolve themselves deterministically.
\end{problem}

Because each elected primary generates \verb+GTID+s with increasing
$term$s, we can predictably get the primary with the lower
$electionTermId$ to step down, and know which updates were accepted by the
deposed primary.  Because all members communicate the highest known
$electionTermId$ throughout the set via heartbeats, we are assured that
the primary with the lower $electionTermId$ will eventually step down,
regardless of who it is connected to.

\begin{problem}
  Elections may take a long time because a member can vote ``yes'' at most
  once every 30 seconds.
\end{problem}

This timer has been removed.  Because Ark is robust enough to handle
multiple primaries and resolves each such case deterministically, we don't
need to have this 30 second timer.

That being said, collisions may still occur.  We may have two candidates,
$B$ and $C$, happen very close in time to each other.

After the first round trip, suppose they both decide on the same
$electionTermId_B = electionTermId_C$.  Both could run authoritative
elections, get a handful of votes, and neither one could get elected.
This is what Raft would call a ``split brain''. In this case, neither
wins, and another election needs to run.

Alternatively, these two elections can happen with different values of
$electionTermId_B < electionTermId_C$, and both might succeed in a short
period of time.  In that case, $B$ is going to become primary and then
almost immediately step down when it sees $electionTermId_C$ has been
voted on.  From a correctness standpoint, that's fine, but it could be
confusing for clients to have to switch primaries too frequently.

These problems exist with the current election protocol.  The symptom in
each protocol is different:
\begin{itemize}
\item Ark can cause two replicas to be primary in quick succession, or
  effectively concurrently for some clients, due to delayed messages.
\item MongoDB's replication incurs a 30 second freeze after unsuccessful
  simultaneous elections.
\end{itemize}
The root cause of these problems is the same: two replicas simultaneously
try to elect themselves.  To mitigate this, the current MongoDB protocol
sleeps for a random amount of time (50-1050ms) before kicking off an
election.  This random delay is used in Raft as well (150-300ms), and is
also used for the same purpose in Ark.

A remaining concern is whether the sleep is sufficient.  Is a random sleep
of up to one second sufficient for replica sets that span across data
centers and may have high ping times?  Raft experiments with a cluster of
5-9 servers with a broadcast time of 15ms determined the recommended sleep
time of 150-300ms.

\section{The solution, in Raft's terms}
\label{solution-raft}

Raft's safety and liveness properties follow directly from five
fundamental properties described in Figure~3 in the Raft paper:
\begin{itemize}
\item {\bf Election Safety}: at most one leader can be elected in a given
  term.
\item {\bf Leader Append-Only}: a leader never overwrites or deletes
  entries in its log; it only appends new entries.
\item {\bf Log Matching}: if two logs contain an entry with the same index
  and term, then the logs are identical in all entries up through the
  given index.
\item {\bf Leader Completeness}: if a log entry is committed in a given
  term, then that entry will be present in the logs of the leaders for all
  higher-numbered terms.
\item {\bf State Machine Safety}: if a server has applied a log entry at a
  given index to its state machine, no other server will ever apply a
  different log entry for the same index.
\end{itemize}

{\bf Election Safety} is provided by the addition and treatment of the
$electionTermId$.  Since each replica can only vote ``yes'' once for each
$electionTermId$, this ensures that in each $electionTermId$ at most one
leader may be elected.

{\bf Leader Append-Only} is an existing property of the oplog and how the
\verb+GTID+ is defined.

{\bf Log Matching} is a consequence of rollback.  Whenever a replica
attempts to sync from another, it asserts the Log Matching property, and
if the source replica's log doesn't contain its oplog as a prefix, the
syncing replica rolls back operations until its oplog matches a prefix of
the source's oplog.  Only after this point may it proceed and copy new
entries.  Multiple concurrent primaries' oplogs may diverge at the end
(and their followers' oplogs would as well), but these entries would have
different terms and therefore would not invalidate the Log Matching
property.

{\bf Leader Completeness} is ensured only for updates acknowledged with
``majority'' Write Concern.  But according to the refusal of replicas to
acknowledge updates after voting for a new primary, Leader Completeness is
ensured for these updates.

{\bf State Machine Safety} makes sense if we consider the long-term state
of the replica set.  In MongoDB/TokuMX, the state machine refers to the
user data collections.  Since updates are applied separately from their
Write Concern acknowledgement, there may be a period of time when an
update is reflected in the state machine of some replicas, before the
operation rolls back.  In this case, the operation's ``index'' within the
log will eventually be taken by a different operation, which would at this
point be applied on other replicas.

This complexity is due to the asynchronous nature of replication in
MongoDB/TokuMX, essentially that application may happen before replication
and then the operation would be rolled back.  However, once a replica
establishes contact with another replica and decides to roll back an
operation (in Raft, this would be when a leader forces its followers to
replace log entries with its own), that operation is rolled back before
any new operations are synced and applied, which maintains logical
separation of the rolled back operation and the legitimately applied
operations.

We needed to tweak the definitions of a few of Raft's fundamental
properties, but not in ways inconsistent with an asynchronous
interpretation of Raft, nor in ways that invalidate its safety and
liveness proofs.  We leave the translation of Raft's safety and liveness
arguments to our asynchronous context as a straightforward but exciting
exercise for the reader.

\section{Conclusions}
\label{conclusions}

We have presented Ark, a new consensus algorithm based on and supporting
the existing MongoDB replication algorithm and semantics, with minimal
changes.  Ark is inspired by Raft and provably provides safety and
liveness according to the same arguments as Raft.

Ark has many of the same understandability properties of Raft.  The oplog
in Ark does not permit holes anywhere in an agreed-upon prefix which
limits the ways in which replicas can diverge from each other.  By virtue
of the MongoDB replication architecture, Ark decouples the mechanisms of
replication and elections, and brings them together just enough to provide
safety.

In contrast with Raft, Ark implements an asynchronous, pull-based
replication model.  This supports a wider range of client semantics that
allows application developers to choose points along a tradeoff between
safety and latency.  In addition, Ark supports different replication
topologies like chained replication and multi-data center replication with
more flexibility than Raft does with its synchronous push model.

While Ark is an implementation of a consensus protocol that works in a
real database system, it is also evidence of the flexibility in the Raft
consensus algorithm.  It was relatively straightforward to tweak Raft in
safe ways to make it fit the MongoDB architecture and programming model,
and we think this is an important feature of Raft.

\section{Future work}
\label{future-work}

We still have some things to think about.  The basics of Ark are correct,
but it's possible there is some tweaking to be done to improve the user
experience.  Likely, most of the below questions can be answered with
experiments and user feedback.

\begin{question}
  Should we veto based on the oplog contents?
\end{question}

In an authoritative election, if the \verb+GTID+ of the candidate $B$'s
last oplog entry is less than the \verb+GTID+ of the voter's, we currently
veto.

We could theoretically vote ``no'' and and still have ``majority'' Write
Concern work properly.  This is a user experience question and not a
correctness question.  Here are reasons why we do this:
\begin{itemize}
\item This should happen in rare circumstances, because speculative
  election does this check and does not proceed if it fails.
\item Users don't just run with ``majority'' Write Concern.  They run with
  other custom Write Concerns to give them different assurances.

  Consider a replica set spread across three data centers, called DC1,
  DC2, and DC3.  Suppose each data center has three members of the replica
  set, so the entire replica set is nine members.  Users may set a custom
  Write Concern that means ``make sure my update has made it to at least
  one other data center''.

  The idea is that the user is assuming either all members in a data
  center remain up or none do, and that the probability of two data
  centers being down is negligible.  Now, suppose the primary is in DC1
  and DC1 goes down.  In this case, if the update only made it to one
  replica in DC2, and that replica just votes ``no'' in the authoritative
  election, then the update got the proper acknowledgement but may be
  rolled back, because the 5 other members that don't have this update may
  elect a different primary.  But with veto power, the update will not be
  rolled back, because the one replica in that did receive the update
  won't let a primary be elected without the update.
\end{itemize}

Essentially, this choice to allow replicas to veto based on the contents
of the oplog lends more consistency to updates that use weaker Write
Concern settings, which seems to be a valuable property for many MongoDB
applications.

\begin{question}
  Should there be some timer so members cannot vote as often as they do?
\end{question}

30 seconds is too long, but should there be something like 1 second?  To
help prevent multiple primaries being elected simultaneously?

In the Raft paper, they spend some time talking about the election
timeout, but in Raft this is the window of time an election stays open to
hear votes.  It is not particularly related to the number of elections a
given node can participate in in a given time slice (apart from this
timeout divided by the number of replicas).

This timer also seems to be valuable in ensuring that the primary doesn't
move too quickly around the set for a client to follow.

\begin{question}
  Is the random sleep of about 1 second before starting an election
  sufficient to keep two competing elections from happening at roughly the
  same time too often?
\end{question}

The sleep time should not be too long, but the range should be
significantly longer than the range of typical message delays in the
cluster.

Should cross-data center replica sets do something different?  One idea is
to pick the sleep time based on ping times we see in heartbeats.

\begin{question}
  Should we make heartbeats more sophisticated?
\end{question}

Currently, all members heartbeat all other members every 2 seconds
regardless of role or topology.  Heartbeat traffic is one of the things
that limits a replica set's size, and for XDCR this could be more traffic
across the WAN than we'd like.

Perhaps each node could heartbeat what it thinks is the primary once a
second, and all other members every 5 seconds?  This could improve failover
time by causing members to notice a down primary faster, and reduce the
overall heartbeat traffic in the cluster.

Another option is to send heartbeats along a different topology than a
complete graph.  Other
systems\cite{Demers:1987:EAR:41840.41841,Ganesh:2001:SPL:648089.747488,epidemic_broadcast_trees}
use rings, trees, or other incomplete graphs to propogate the same
information being exchanged here with heartbeats.

\begin{question}
  How should we handle dynamic cluster membership?
\end{question}

Raft has an elegant solution to dynamic membership: membership changes are
recorded in the log stream, and the cluster uses an intermediate stage
where replicas are aware of both configurations temporarily, until they
have agreed on the new configuration.

MongoDB's existing configuration management is currently stored outside of
the log, and the in-memory representation of configuration is not well
suited to maintaining the intermediate dual-configuration state used by
Raft.  It is still an open question as to the right way to adapt a safe
dynamic membership protocol to the existing MongoDB architecture.

Storing configuration changes in the log seems fairly straightforward and
appropriate, but managing the dual-configuration state properly seems like
a challenge.

\bibliographystyle{plain}
\bibliography{ark}

\end{document}